\documentclass[a4paper,fleqn]{cas-sc}

\usepackage{graphicx,subfigure}
\usepackage[authoryear]{natbib}

\begin{document}

\title{Construct a realistic stellar model with polytropic relation}
\author[1,2]{Xing Wei \footnote{xingwei@bnu.edu.cn}}[orcid=0000-0002-3641-6732]
\affiliation[1]{Institute for Frontier in Astronomy and Astrophysics; Beijing Normal University; China}
\affiliation[2]{Department of Astronomy; Beijing Normal University; China}
\shorttitle{Realistic stellar model}
\shortauthors{Xing Wei}

\begin{abstract}
The aim of this study is to construct a simple stellar model with non-uniform polytropic index. We find that the Emden equation cannot deal with the polytrope gas sphere with non-uniform polytropic index in a real star, and then we construct a realistic stellar model. The key point is that we should solve the two independent equations for density and pressure due to the essence of polytropic relation, but not the Emden equation which combines the hydrostatic balance and polytropic relation. We take the Sun for a computational example to find that this simple model yields quite a good result compared to the MESA code. The advantage of this simple model lies in its much simpler equation of state than that in the standard stellar model.
\end{abstract}

\maketitle

\section{Introduction}
Polytrope gas sphere was once used in the early study of stellar structure \citep{chandrasekhar1939}. With the development of modern computers, researchers tend to solve much more complex stellar structure equations \citep{kippenhahn1990} with sophisticated numerical codes, e.g. the MESA code \citep{Paxton2011, Paxton2013, Paxton2015, Paxton2018, Paxton2019}. However, polytrope model has its advantage, namely it has very simple equation of state but in the standard stellar model the complex tables for equation of state and opacity are used. In some stars, the equation of state is unknown and the polytropic relation is often used. Or when we calculate stellar oscillations with very fast rotation or strong magnetic field, because the problem becomes two dimensional (Coriolis or Lorentz force couples different degrees of spherical harmonics), we have to develop a new code other than the MESA code. In these situations, it is better to adopt a simple equation of state, namely the polytropic relation.

\begin{figure}
\centering
\includegraphics[scale=0.57]{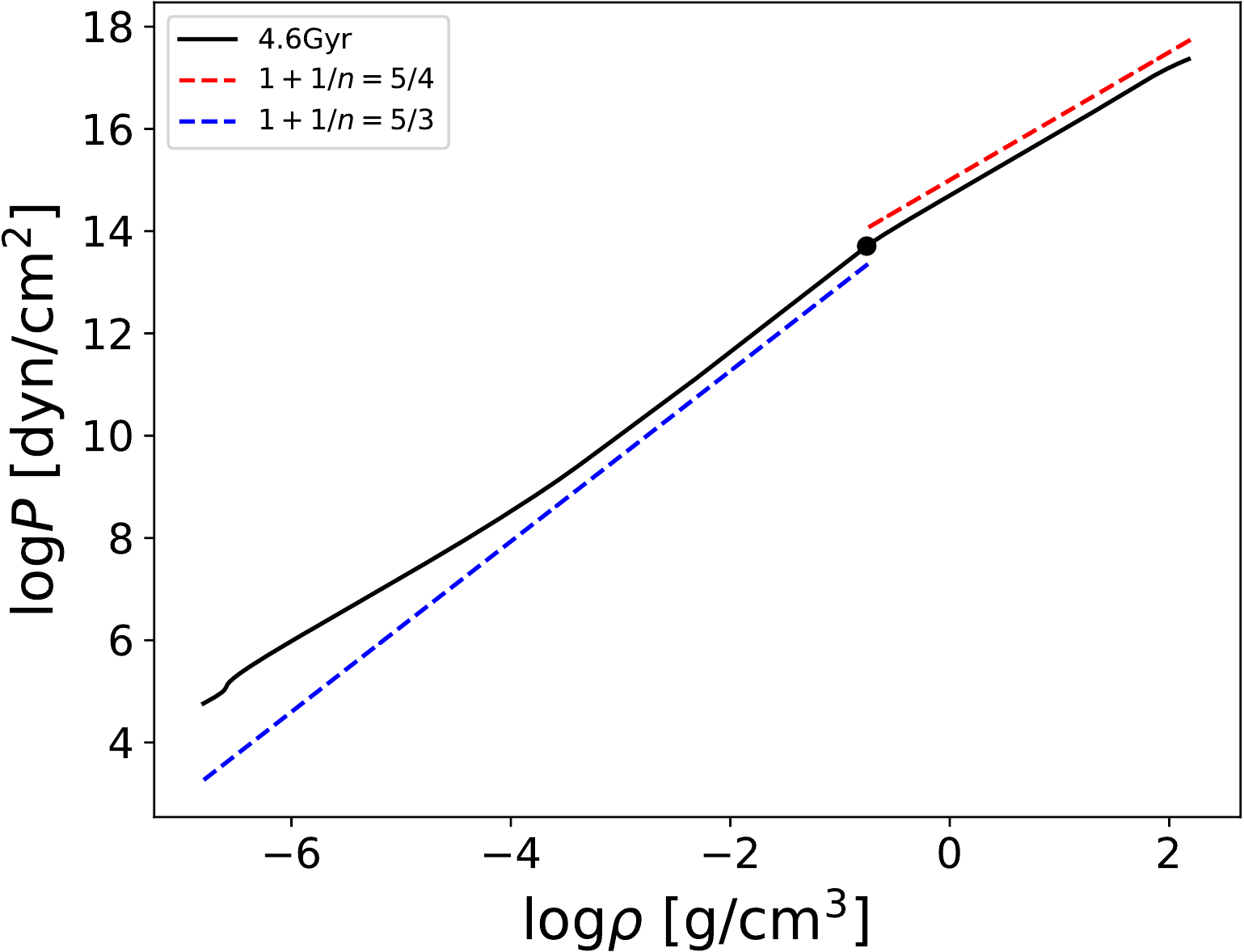}
\caption{Pressure-density relation for the present Sun. Black line denotes the result with MESA code, and the red and blue lines denote the fitting results for different polytropic index. The dot denotes the radiation-convection boundary (RCB). }\label{fig1}
\end{figure}

In the first place, we have a look at the result of the MESA code for the present Sun. Figure \ref{fig1} shows the pressure-density relation and the dot denotes the radiation-convection boundary (RCB), i.e. the location of tachocline. Fitting the curve gives the non-uniform polytropic index, namely $n=1.5$ in convection zone and $n=4$ in radiation zone. $n=1.5$ is for adiabatic process in convection zone. The fact that $n=4$ but not 3 in radiation zone arises from opacity. The temperature gradient $\nabla_{\rm rad}=(3/16\pi acG)(\kappa L_rP/M_rT^4)$ keeps constant. Inserting the polytropic relation $P\propto\rho^{1+1/n}$, the equation of state for ideal gas $P=(\mathcal{R}/\mu)\rho T$ and the Kramers opacity $\kappa\propto\rho T^{-3.5}$, we readily obtain $n=3.25$ (not exactly 4 because Kramers opacity is inaccurate). Such situation of non-uniform polytropic index widely exists in stars or planets \citep{Basillais2021}. Next we will use the polytropic relation with non-uniform $n$ to construct the stellar model so as to avoid the complex standard stellar structure equations and the complex tables for equation of state and opacity.

\section{Method and results}
\begin{figure}
\centering
\includegraphics[scale=0.5]{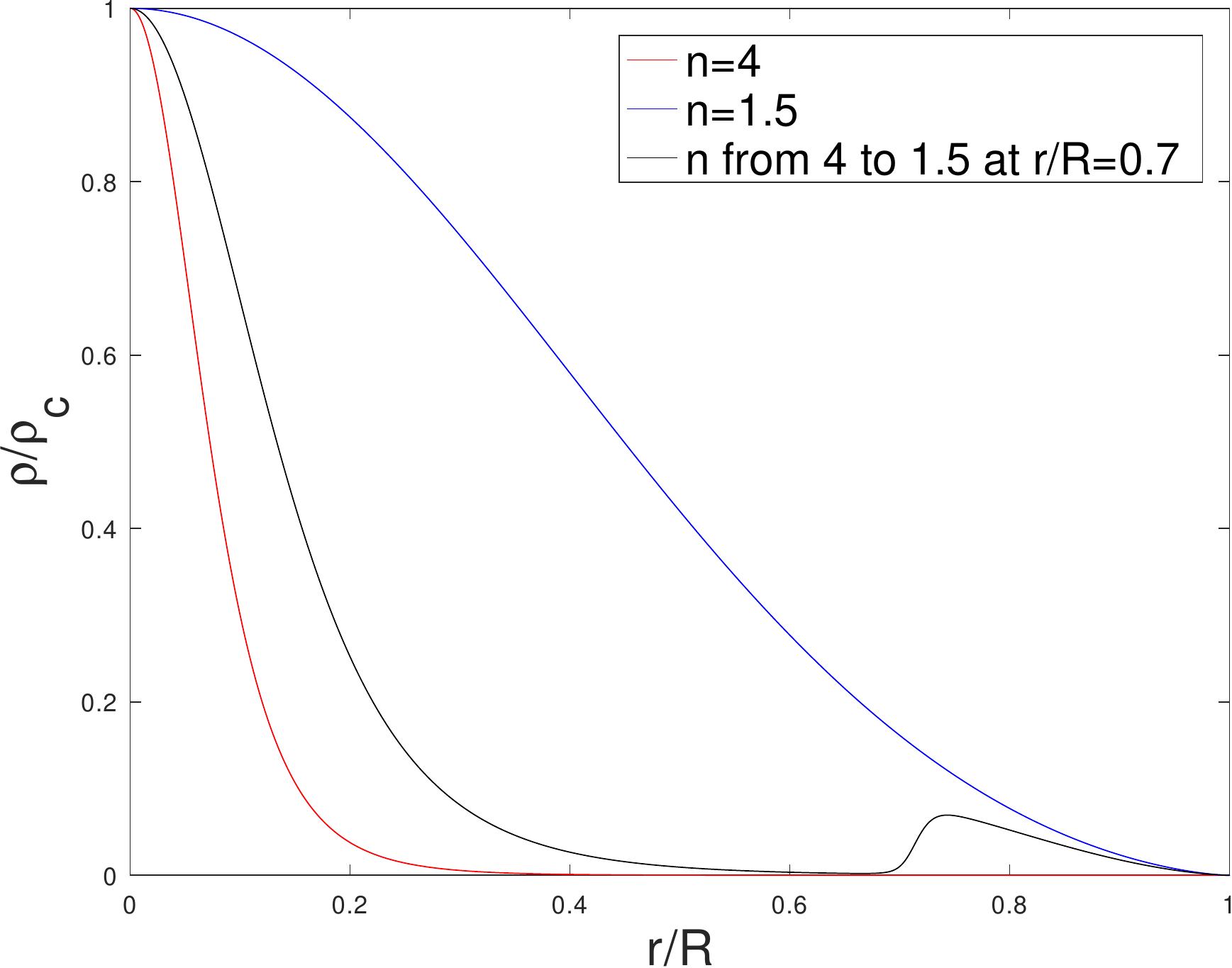}
\caption{The three solutions to the Emden equation with different polytropic index $n$. Horizontal axis denotes normalized radius and vertical axis denotes normalized density.}\label{fig2}
\end{figure}

Firstly we try to solve the Emden equation with non-uniform polytropic index $n$. We give the transition of $n$ from 4 to 1.5 at RCB and smooth the transition with hyperbolic function $tanh$,
\begin{equation}\label{n}
n=2.75-1.25\tanh\frac{r/R-r_0/R}{\sigma}
\end{equation}
where $r_0/R$ denotes RCB, given to be 0.7 for the present Sun, and $\sigma$ denotes the width of transition layer, given to be 0.001 in our computation. Figure \ref{fig2} shows the three solutions to the Emden equation with, respectively, $n=4$, $n=1.5$ and $n$ given by \eqref{n}. The solution with \eqref{n} appears a hump at $r_0/R=0.7$ to connect the inner solution with $n=4$ and the outer solution with $n=1.5$. This hump is definitely unrealistic in physics, i.e. density should not increase with radius. After many numerical tests we find the two reasons for this hump. One is that the Emden equation does not involve the radial derivative of $n$, and the other, more importantly, is the expression of polytropic relation that the Emden equation adopts. To derive the Emden equation \citep{kippenhahn1990} we use the polytropic relation $P=K\rho^{1+1/n}$ that brings the numerical difficulty with non-uniform $n$, no matter how accurate the numerical scheme is chosen. To avoid this difficulty, we should use the more essential expression $d\log P=(1+1/n)d\log\rho$.

We now design our new approach. We write down the hydrostatic balance and polytropic relation with its essential expression,
\begin{align}
&\frac{dP}{dr}=-\rho g=-\rho\frac{4\pi G}{r^2}\int_0^r \rho r'^2dr', \\
&d\log P=(1+1/n)d\log\rho.  
\end{align}
We next normalize these two equations. Density $\rho$ is normalized with central density $\rho_c$, pressure $P$ with central pressure $P_c$, and radius $r$ with $r_0$ which will be given later. We use tilde to denote the normalized quantities so that the above two equations become
\begin{align}
&\frac{d\tilde P}{d\tilde r}+\frac{4\pi G\rho_c^2r_0^2}{P_c}\frac{\tilde\rho}{\tilde r^2}\int_0^{\tilde r}\tilde\rho\tilde r'^2d\tilde r'=0, \label{hydro}\\
&d\log \tilde P=(1+1/n)d\log\tilde\rho. \label{polytropic}
\end{align}
By setting the coefficient $4\pi G\rho_c^2r_0^2/P_c=1$ in \eqref{hydro} we immediately obtain the radius unit $r_0=\sqrt{P_c/4\pi G}/\rho_c$. 

We then choose the first-order difference method to numerically solve \eqref{hydro} and \eqref{polytropic}. The numerical scheme is as follows
\begin{align}
\tilde P_{i+1}&=\tilde P_i-\frac{\tilde\rho_i\Delta\tilde r}{\tilde r_i^2}\int_0^{\tilde r_i}\tilde\rho\tilde r'^2d\tilde r', \label{hydro-num}\\
\log\tilde\rho_{i+1}&=\log\tilde\rho_i+\frac{n}{n+1}(\log\tilde P_{i+1}-\log\tilde P_i), \label{polytropic-num}
\end{align}
where $\Delta\tilde r$ is the integration step. The integral $\int_0^{\tilde r_i}\tilde\rho\tilde r'^2d\tilde r'$ is evaluated by the first-order trapezoidal rule. The boundary conditions at the center are $\tilde\rho=1$ and $\tilde P=1$. Thus, we integrate from the center, by \eqref{hydro-num} we obtain pressure $\tilde P_{i+1}$ and then by \eqref{polytropic-num} we obtain density $\tilde\rho_{i+1}$. It should be noted that in the above numerical scheme the radial derivative of $n$ is already involved. Although we use the first-order scheme, we find that the solution is fairly good when the integration points are sufficient.

\begin{figure}
\centering
\includegraphics[scale=0.5]{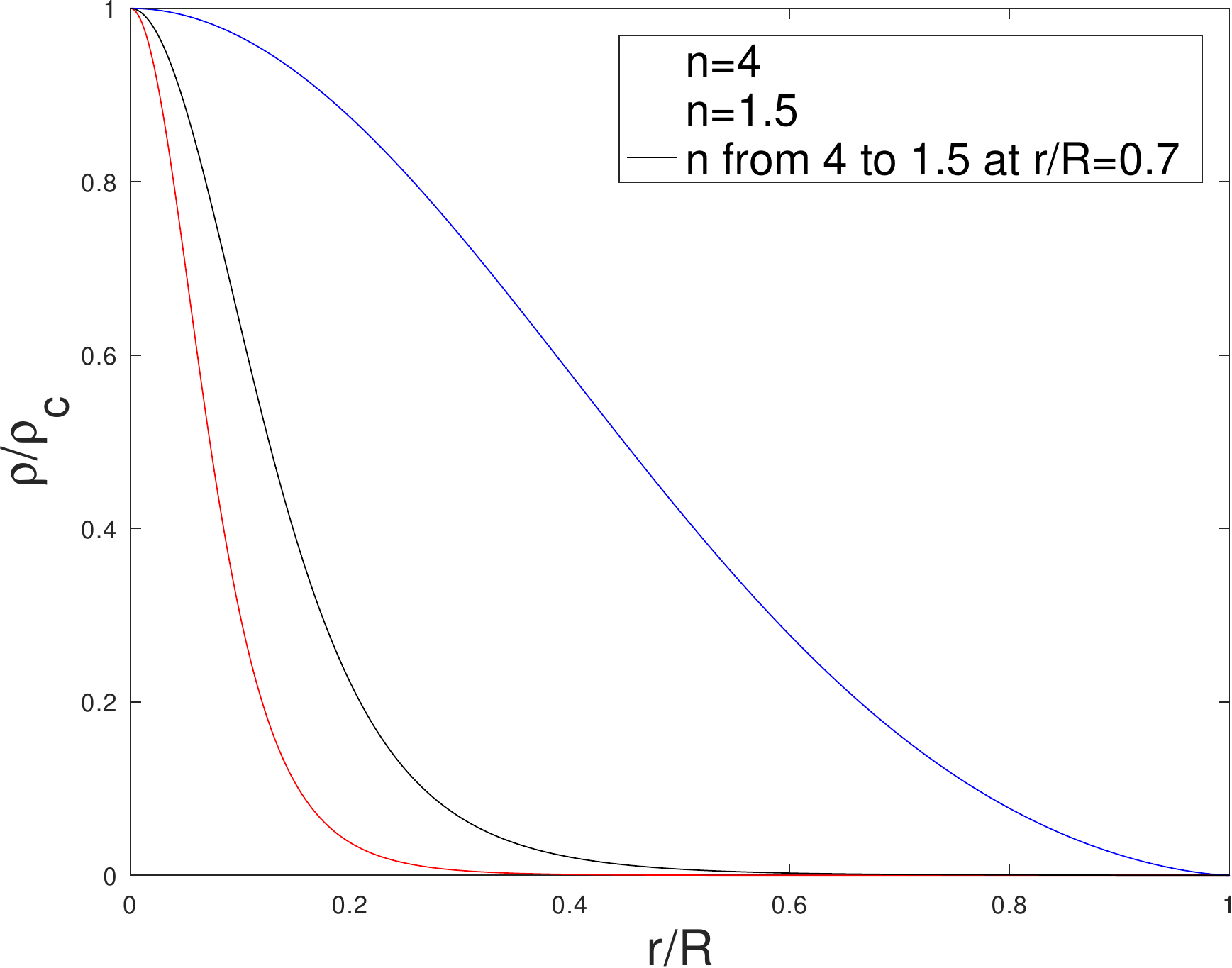}
\caption{Similar to Figure \ref{fig2} but the black curve is our result with Eqs. \eqref{hydro} and \eqref{polytropic}.}\label{fig3}
\end{figure}

Figure \ref{fig3} shows our result with $n$ given by \eqref{n}. Compared to Figure \ref{fig2}, we find that the hump disappears. The major difference of our method from the Emden equation is that we solve density and pressure with the two independent equations, whereas in the Emden equation pressure is eliminated through $P=K\rho^{1+1/n}$ but only density is kept as variable. As we mentioned before, the essence of polytropic relation is $d\log P=(1+1/n)d\log\rho$ but not $P=K\rho^{1+1/n}$. This is the key reason that this hump appears in the Emden equation with non-uniform $n$.

\begin{figure}
\centering
\includegraphics[scale=0.5]{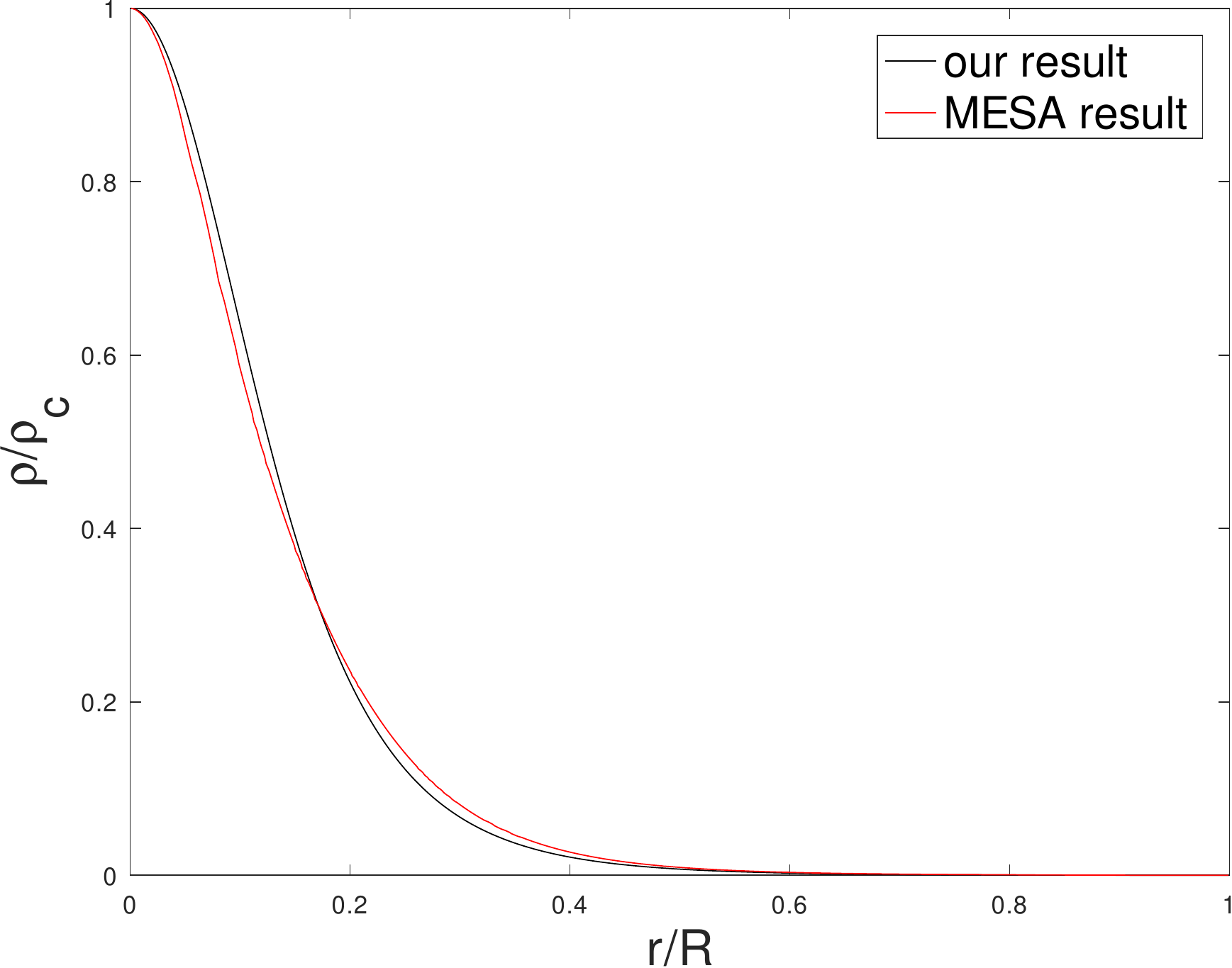}
\caption{Comparision between our result with Eqs. \eqref{hydro} and \eqref{polytropic} and the MESA result.}\label{fig4}
\end{figure}

We compare our result to the MESA result. Figure \ref{fig4} shows the density profile of the two results, and our result fits well with the MESA result. We also compare the physical quantities at the center. The stellar total mass can be calculated as 
\begin{equation}\label{mass}
M=4\pi\left(\frac{R}{\tilde r_{max}}\right)^3\rho_c\int_0^{\tilde r_{max}}\tilde\rho\tilde r'^2d\tilde r'
\end{equation}
where $\tilde r_{max}$ is the dimensionless stellar radius at which $\tilde\rho\approx 0$ (in the computation we set the lower limit $10^{-7}$). On the other hand, $M=(4/3)\pi R^3\bar\rho$ where $\bar\rho$ is mean density. By \eqref{mass} we obtain $\bar\rho/\rho_c=3\int_0^{\tilde r_{max}}\tilde\rho\tilde r'^2d\tilde r'/\tilde r_{max}^3$ such that we can find $\rho_c$. Using $R=r_0\tilde r_{max}$ and the radius unit $r_0=\sqrt{P_c/4\pi G}/\rho_c$ we can find $P_c$. With the equation of state for ideal gas $P=(\mathcal{R}/\mu)\rho T$ we can find $T_c$. Table \ref{table} shows the physical quantities at the center of the Sun in our result and in the MESA result as well as the relative error. Although our density profile fits the MESA result very well (Figure \ref{fig3}), the relative error is not very good, especially the relative error of central pressure reaches 0.24. This is reasonable, because the MESA code solves the complex stellar structure equations with the complex tables for equation of state and opacity whereas we use a very simple polytropic relation.

\begin{table}
\centering
\begin{tabular}{|c|c|c|c|c|c|}
\hline
& MESA result & our result & relative error \\ \hline
$\rho_c ~(g/cm^3)$ & 151 & 166 & 0.10  \\ \hline
$P_c ~(dyn/cm^2)$ & 2.32E17 & 3.04E17 & 0.24  \\ \hline
$T_c ~(K)$ & 1.56E7 & 1.35E7 & 0.13 \\ \hline
\end{tabular}
\caption{Central quantities in our result and the MESA result.}\label{table}
\end{table}

\section{Conclusion}
Due to the essence of polytropic relation, i.e. $d\log P=(1+1/n)d\log\rho$ but not $P=K\rho^{1+1/n}$, for non-uniform polytropic index, we need to solve the two independent equations for density and pressure but not the combined Emden equation. This simple model with polytropic relation can be used to construct the realistic stellar or planetary structure. We compared this model to the solar structure obtained by the MESA code and the result is fairly good. Readers may try higher-order numerical schemes (we use only first-order) to improve the accuracy of computational result, but should keep in mind that the two equations cannot be combined.

\section*{Acknowledgments}
I thank Tao Cai for their helpful discussions. Qiang Hou provides the MESA result. This work is supported by National Natural Science Foundation of China (11872246, 12041301).

\bibliographystyle{cas-model2-names}
\bibliography{paper}
\end{document}